\documentclass[aps,prl,epsfigure,twocolumn,superscriptaddress]{revtex4-2}
\usepackage[colorlinks=true,linkcolor=blue,urlcolor=blue,citecolor=blue,pdfusetitle]{hyperref}
\usepackage[utf8]{inputenc}
\usepackage[english]{babel}
\usepackage{amsmath}
\usepackage[caption = false]{subfig}
\usepackage{graphicx,epstopdf}
\usepackage{blindtext}
\usepackage[table,xcdraw]{xcolor}
\usepackage{lipsum}
\usepackage{amsfonts}
\usepackage{bbm}
\usepackage{amssymb}
\usepackage{enumerate}
\usepackage{color}
\usepackage{latexsym}
\usepackage{times,txfonts}

\newcommand{\Dcal}{\mathcal{D}}

\newcommand{\1}{\mathbbm{1}}

\newcommand{\ket}[1]{| #1 \rangle}
\newcommand{\bra}[1]{\langle #1 |}

\newcommand{\SubFig}[2]{\ref{#1}{\color{blue}#2}}

\definecolor{MyGreen}{RGB}{0, 179, 134}
\definecolor{MyRed}{RGB}{255, 102, 102}

\newcommand{\UFSCar}{Departamento de Física, Universidade Federal de São Carlos, Rodovia Washington Luís, km 235 - SP-310, 13565-905 São Carlos, SP, Brazil}
\newcommand{\CSIC}{Instituto de Física Fundamental (IFF), Consejo Superior de Investigaciones Científicas (CSIC), Calle Serrano 113b, 28006 Madrid, Spain}

\usepackage{orcidlink}

\begin{document}
	
	\title{Interference between non-overlapping waves}
	
	\author{Alan C. Santos \orcidlink{0000-0002-6989-7958}}%
\email{ac\_santos@iff.csic.es}
\affiliation{\CSIC}

\author{Celso J. Villas-Boas \orcidlink{0000-0001-5622-786X}}%
\email{celsovb@df.ufscar.br}
\affiliation{\UFSCar}

\begin{abstract}
In classical mechanics and electromagnetism, interference occurs when two or more waves overlap at the same point in spacetime. However, the advent of quantum electrodynamics (QED) and its remarkable success in describing light–matter interactions at the microscopic level invites us to reconsider whether interference-like effects could arise even when the waves do not physically overlap. In this work, we extend the notion of wave interference to a novel and unconventional regime. Building upon the fundamental description of interference in terms of the interaction with the observer [\href{https://journals.aps.org/prl/abstract/10.1103/PhysRevLett.134.133603}{Phys. Rev. Lett. \textbf{134}, 133603 (2025)}], we demonstrate that interference-like phenomena can emerge when two independent fields interact with a single detector at different locations in Minkowski space. We begin by developing a theoretical model in which a spatially extended atom simultaneously couples to two distant fields. We then propose an experimentally feasible implementation using superconducting circuits, where a giant artificial atom interacts with two independent resonators. Our findings open new directions for exploring interference in quantum systems and suggest new possibilities for optical quantum technologies, including the realization of atom-transparent devices controlled by spatially separated laser fields.

\end{abstract}

\maketitle

\textit{Introduction.}-- In classical theory, the phenomena of wave interference and its consequences are primarily described by the summation of quantities---mathematical functions---carrying information about the properties of such waves. In this picture, the absence of observables or detection apparatus is not relevant to explain the nature of interference~\cite{huygens1912treatise, darrigol2012history}. On the other hand, with the advent of quantum theory, a universe of new possibilities has led to the accurate description of unexplained phenomena and processes occurring in nature. In particular, the Dirac's~\cite{dirac:27} and Fermi's~\cite{fermi:32} quantum theories of radiation, and the quantum theory of light-matter interaction~\cite{greiner2008,sheremet:23} opened a broad avenue to the prediction of new effects and development of theories to explain how the microscopical matter behaves when interacting with light. As consequence, the reinterpretation of the phenomenon of interference can be done by introducing microscopic models of the interaction between the different field modes and the measurement apparatus---e.g., photon detectors can be modeled as a two-level system~\cite{Glauber1963, glauber2007quantum, Villas-Boas2025}.

Regardless of the theory used to describe the interference, it is widely presumed that it only occurs when the waves overlap at a given point in the Minkowski spacetime. However, quantum theory allows us to rethink the origins of classical interference, where we can reinterpret and explain wave interference by considering the corpuscular character of both matter and light through the existence of collective effects. Through this corpuscular interpretation of light-matter interaction, we can predict light particle states that couple to matter, known as bright states, and states that do not couple, referred to as dark states~\cite{Villas-Boas2025}. These states are related to constructive and destructive interference, respectively, capable of explaining the double slit experiment~\cite{Villas-Boas2025}, and propose solutions to open problems like the mode-locked pulsed lasers~\cite{diniz2024}. In addition, the understanding of these dark states for light finds further applications in quantum technologies, like new optical devices for light control~\cite{Luiz:25}.

In this work, we use the conceptual perspective of \textit{collective light states} that can either couple (bright) or does not couple (dark) with the detector recently proposed in~\cite{Villas-Boas2025}, to incite the discussion around the hypothesis that the coincidence of modes in the spacetime is not required for light interference. To this end, we demonstrate that it is possible to observe interference-like effects between two waves interacting with a common observer, even without any spatial overlap. In particular, we propose an experimentally feasible example where an artificial superconducting giant atom interacts with two single-frequency superconducting resonators.

\begin{figure*}[t!]
\centering
\includegraphics[width=\linewidth]{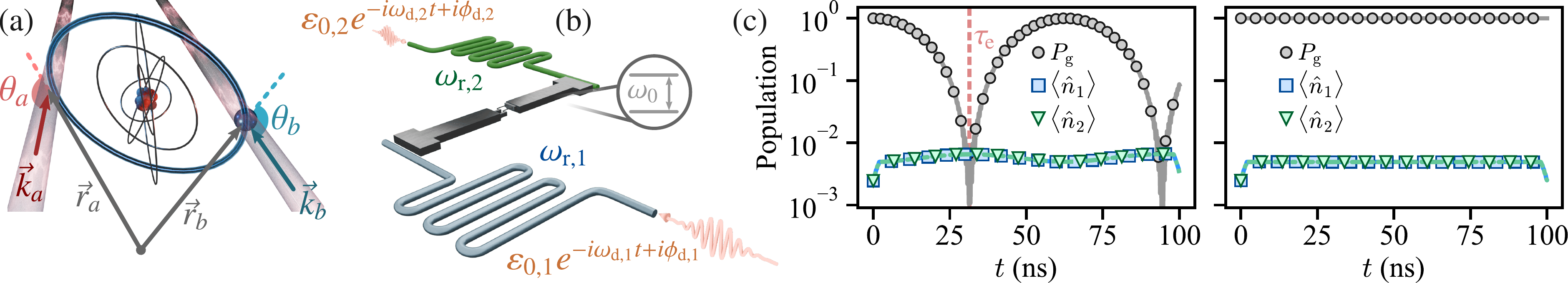}
\caption{(a) Representation of a spatially extended atom interacting with two independent fields at positions $\vec{r}_{a}$ and $\vec{r}_{b}$. There is no spatial overlap between the fields. This system can be efficiently simulated in superconducting quantum devices, as shown (b). Two off-resonant resonators ($\omega_{\mathrm{r},1}\neq \omega_{\mathrm{r},2}$) coupled to a giant artificial atom in such way that no direct parasitic or collateral interaction takes place. Local oscillating fields are applied in each resonator to excite the atom in the qubit subspace. (c) The qubit population to out-of-phase (left) and in-phase drives applied to each resonator. We show the average photon population for each mode. The parameters used in these figures are discussed in the text.}
\label{Fig:Scheme}
\end{figure*}

\emph{Atom driven by non-overlapping classical fields.}-- Here we consider the following scenario: A single atom is allowed to interact with two \textit{localized} distinct fields, as sketched in Fig.~\SubFig{Fig:Scheme}{a}. The two modes are described by annihilation (creation) operators $a$ and $b$ ($a^{\dagger}$ and $b^{\dagger}$). For simplicity, we assume both modes have the same frequency $\omega$---the results are also applicable to non-resonant modes, as we shall see. We consider these modes to interact with the giant atom, modeled as a two-level system with ground state $\ket{g}$, excited state $\ket{e}$, and transition frequency $\omega_0$. Thus, given the static electric field vector operator for each mode as $\hat{E}_{a} \propto \hat{a} e^{i\vec{k}_a \cdot \vec{r}_a} + h.c.$ and $\hat{E}_{b} \propto \hat{b} e^{i\vec{k}_b \cdot \vec{r}_b} + h.c.$, and assuming the emitter with dipole momentum $\hat{d}$, the total Hamiltonian of the system is then given by ($\hbar = 1$) $H = \hat{H}_0 + \hat{H}_I$, where $\hat{H}_0 = \omega \hat{a}^{\dagger}\hat{a} + \omega \hat{b}^{\dagger}\hat{b} + \omega_0 \hat{\sigma}_z/2$, and
\begin{equation}
\hat{H}_I 
=-\left(\hat{E}_{a} + \hat{E}_{b}\right)\cdot \hat{d} \overset{\text{\tiny{RWA}}}{=} g\left(\hat{a} e^{i\vec{k}_a \cdot \vec{r}_a} + \hat{b} e^{i\vec{k}_b \cdot \vec{r}_b} \right)\hat{\sigma}_{+} + h.c., \label{Eq:Hamil}
\end{equation}
with $\hat{\sigma}_z = \ket{e}\bra{e} - \ket{g}\bra{g}$, $\hat{\sigma}_{+} = \ket{e}\bra{g} = \hat{\sigma}_-^{\dagger}$, and $g$ being the atom-field coupling strength, assumed to be the same for both modes, for simplicity. Here, $\vec{k}_{x}$ and $\vec{r}_{x}$, with $x = a, b$, denote the wave vector of each mode and the corresponding position where each mode interacts with the giant atom, respectively. The term \textit{h.c.} stands for the Hermitian conjugate. In this equation, RWA refers to the Rotating Wave Approximation, which is valid when the atom-field coupling is much weaker than the mode and atomic transition frequencies~\cite{scully1997quantum,walls2008quantum}.

It is convenient to rewrite Eq.~\eqref{Eq:Hamil} in terms of a new collective mode, $\hat{C}(\vec{r}_{a},\vec{r}_{b})$, as $\hat{H}_I = g \sqrt{2} \hat{C}(\vec{r}_{a},\vec{r}_{b})\hat{\sigma}_{+} + h.c.$, where the collective operator is
\begin{equation}
\hat{C}(\vec{r}_{a},\vec{r}_{b}) = \frac{1}{\sqrt{2}} \left(\hat{a} e^{i\vec{k}_{a} \cdot \vec{r}_a} + \hat{b} e^{i\vec{k}_b \cdot \vec{r}_b}\right) .
\end{equation}

We understand the physical interpretation of this collective operator as a result of the linear combination of two independent and spatially delocalized modes. As a result, the total number of photons in this collective mode is
\begin{align}
\hat{N}_{\mathrm{col}} &= \hat{C}^{\dagger}(\vec{r}_{a},\vec{r}_{b})\hat{C}(\vec{r}_{a},\vec{r}_{b})
\nonumber \\
&= \frac{1}{2} \left[ \hat{N}_{\mathrm{total}} + \hat{b}\hat{a}^{\dagger} e^{i(\vec{k}_{b} \cdot \vec{r}_{b}-\vec{k}_{a} \cdot \vec{r}_{a})} + \hat{b}^{\dagger}\hat{a} e^{-i(\vec{k}_{b} \cdot \vec{r}_{b}-\vec{k}_{a} \cdot \vec{r}_{a})}\right] ,
\end{align}
which allows us to identify that $\hat{N}_{\mathrm{col}} \neq \hat{N}_{\mathrm{total}} = \hat{a}^{\dagger}\hat{a} + \hat{b}^{\dagger}\hat{b}$. This mathematical analysis leads to a first counterintuitive result of our work. The presence of photons in the individual modes $\hat{a}$ and $\hat{b}$ is not sufficient to ensure energy transfer between the field and the atom, since the absence of excitation in the collective mode $\hat{C}(\vec{r}_a, \vec{r}_b)$ prevents the interaction to take place. Moreover, as a second counterintuitive result, the atom will feel an interference-like effect when interacting with these two independent modes, even when the atom is driven by spatially separated fields. It can be shown by considering the mode states as $\ket{\Psi} = \ket{\psi_{a}}\otimes \ket{\psi_{b}}$, and evaluating $\hat{N}_{\mathrm{col}}$ as
\begin{align}
\langle \hat{N}_{\mathrm{col}} \rangle_{\Psi} = \frac{N_{\mathrm{total}}}{2} + \Re[\langle\hat{b}\rangle_{\psi_{b}}\langle\hat{a}^{\dagger}\rangle_{\psi_{b}} e^{i(\vec{k}_{b} \cdot \vec{r}_{b}-\vec{k}_{a} \cdot \vec{r}_{a})}] \leq N_{\mathrm{total}} ,
\end{align}
with $N_{\mathrm{total}} = \langle\hat{N}_{\mathrm{total}}\rangle_{\Psi}$ the total number of photons in the bare modes. The above equation constitutes clear evidence of interference-like effects in the collective mode $\hat{C}(\vec{r}_{a},\vec{r}_{b})$, and it is valid even for states of the fields without any classical or quantum correlation. As a notable remark, although this result does not suggest any real interaction between the modes, as it is excluded by definition in our model, this effect is crucial for the effective field interacting with the atom.

To demonstrate this result, consider that each mode is \textit{independently} in a coherent state, such that the joint state for the modes is $\ket{\alpha,\beta} = \ket{\alpha}_{a} \otimes \ket{\beta}_{b}$. In the classical regime of the coherent fields, it is possible to show that the dynamics of the emitter can be described by the resulting semi-classical Hamiltonian~\cite{scully1997quantum,walls2008quantum} 
\begin{align}
\hat{H}_{\mathrm{res}} = g  \left[ \alpha e^{i\vec{k}_a \cdot \vec{r}_a} + \beta e^{i\vec{k}_b \cdot \vec{r}_b } \right]\hat{\sigma}_{+} + h.c. \label{Eq:H_res}
\end{align}


As a result, we understand the Hamiltonian $\hat{H}_{\mathrm{res}}$ as the result of a superposition between the two \textit{non-overlapping} modes. To support this claim, notice that $\hat{H}_{\mathrm{res}}$ is similar to the driving Hamiltonian of an effective mode with resulting electric field with amplitude $E_\mathrm{res} \propto ( \alpha e^{i\vec{k}_a \cdot \vec{r}_a} + \beta e^{i\vec{k}_b \cdot \vec{r}_b } )$. 
Furthermore, this analogy becomes even more evident when we take into account possible destructive and constructive interference-like effects. 

In fact, for simplicity we set $|\vec{k}_b|=|\vec{k}_a|=|\vec{k}_0| = 2\pi/\lambda$, such that $\vec{k}_x \cdot \vec{r}_x = 2\pi|\vec{r}_x|\cos(\theta_{x})/\lambda$, with the geometrical meaning of the angle $\theta_{x}$ shown in Fig.~\SubFig{Fig:Scheme}{a}. Therefore, the amplitude of the resulting electric field $E_\mathrm{res}$ can be obtained as
\begin{align}
|E_\mathrm{res}| \propto \left\vert\alpha  + \beta e^{i\frac{2\pi}{\lambda} \big[|\vec{r}_b|\cos(\theta_{b}) - |\vec{r}_a|\cos(\theta_{a})\big] }\right\vert .
\end{align}

Thus, we observe that by adjusting the interaction positions such that $|\vec{r}_b|\cos(\theta_{b}) - |\vec{r}_a|\cos(\theta_{a}) = n \lambda$, with $n \in \mathbbm{Z}$, the above equation yields $|E_\mathrm{res}| \propto |\alpha + \beta|$, which corresponds to constructive interference whenever $\mathrm{sign}(\alpha)=\mathrm{sign}(\beta)$, and destructive interference if $\mathrm{sign}(\alpha)=-\mathrm{sign}(\beta)$. In the case of constructive interference, we observe that the rate of energy exchange between the modes and the atom is amplified. In particular, it is worth mentioning that for identical and for out-of-phase quasi-classical states, i.e., $\alpha=-\beta$, we obtain $|E_\mathrm{res}| = 0$, corresponding to perfectly destructive interference, and the fields are unable to excite the atom. 

The non-intuitive detail of this result is the \textit{delocalized} nature of the modes. It implies that interference occurs within the atom itself even for two waves without any spatial overlap, shedding new light on the fundamental nature of interference phenomena. In the following, we analyze this in greater detail by quantifying the photon exchange with a giant atom under constructive and destructive interference conditions. Therefore, it differs from the classical description of interference, where the superposition is ``witnessed'' through the resultant phasor at a given point in the Minkowski space. However, it is possible to think of an experimental setup in which this delocalized interference can be witnessed through a resultant phasor. Now, we analyze an experimentally feasible proposal that allows for the investigation of such concepts experimentally in the context of giant atoms in superconducting circuits.

\emph{Experimental proposal.--} Here we propose a scheme to verify the previous results using artificial superconducting giant emitters coupled to single-mode resonators. The system considered here is as sketched in Fig.~\SubFig{Fig:Scheme}{b}. A single superconducting artificial giant atom interacts with two independent (quasi-1D) waveguides, A and B, in two different configurations to exploit its properties as a giant atom. A similar system has been used with a $\Delta$-type giant atom in Ref.~\cite{Chen:22}. In any case, the bare system is described by the Hamiltonian
\begin{eqnarray}
\hat{H}_{0} = \hbar \sum\nolimits_{k=1}^{2}\omega_{\mathrm{r},k} \hat{r}_{k}^{\dagger}\hat{r}_{k} + \hbar \omega_{0}\hat{q}^{\dagger}\hat{q} + \hbar \frac{\alpha}{2}\hat{q}^{\dagger}\hat{q}^{\dagger}\hat{q}\hat{q} ,
\end{eqnarray}
where the first term refers to the resonators' energies, with frequencies $\omega_{\mathrm{r},k}$ and annihilation (creation) operators $\hat{r}_{k}$ ($\hat{r}_{k}^{\dagger}$), and the second term is the atom Hamiltonian, with natural frequency $\omega_{0}$, anharmonicity $\alpha$, and annihilation (creation) operator $\hat{q}$ ($\hat{q}^{\dagger}$) (to be approximated by to a two-level system in the limit of large anharmonicity). We assume that an oscillating drive (frequency $\omega_{d,k}$, relative phase $\phi_{d,k}$, and strength $\varepsilon_{d,k}$) is applied to the system through the resonators, whose Hamiltonian reads~\cite{Blais:04,Santos:23a,Blais:07,Mitchell:21}
\begin{eqnarray}
\hat{H}_{\mathrm{d}}(t) = \sum\nolimits_{k=1}^{2}\hbar\varepsilon_{\mathrm{d},k}\left( \hat{r}_{k}^{\dagger}e^{-i\omega_{\mathrm{d},k} t + i\phi_{\mathrm{d},k}} + \hat{r}_{k}e^{i\omega_{\mathrm{d},k} t - i \phi_{\mathrm{d},k}}\right) .
\end{eqnarray}	
These fields correspond to spatially separated resonators, so we can describe the driving fields using different modes. Finally, the interaction between the atom and the resonators is described by the Hamiltonian
\begin{eqnarray}
\hat{H}_{\mathrm{c}} = \hbar \sum\nolimits_{k=1}^{2} g_{k} \left(\hat{q}^{\dagger}\hat{r}_{k} + \hat{q}\hat{r}_{k}^{\dagger}\right)  ,
\end{eqnarray}
with $g_k$ denoting the coupling between the atom and the $k$-th resonator.

The exact dynamics of the system is then governed by the time-dependent Schrödinger equation
\begin{align}
i\hbar\ket{\dot{\Psi}(t)} = \left[ \hat{H}_{0} +\hat{H}_{\mathrm{d}}(t)+\hat{H}_{\mathrm{c}} \right]\ket{\Psi(t)} . \label{Eq:Dynamics}
\end{align}
However, this complex dynamic including two resonators and the atom can be efficiently reduced to a dynamical equation involving only the atom for certain particular dynamics. For instance, here we assume that the frequency of each waveguide is detuned from the giant atom frequency transition according to the dispersive regime of interaction, $|\Delta_{k}| = |\omega_{\mathrm{r},k} - \omega_{0}| \gg |g_{k}|$. In this limit, we guarantee that correlations between the atom and waveguide are suppressed and the total state can be approximated by $\ket{\Psi(t)} = \ket{\psi(t)}_{\mathrm{atom}}\ket{\chi_{1}(t)}_{r_{1}}\ket{\chi_{2}(t)}_{r_{2}}$. This allows us to simplify the description of the system through a transformation in the Schrödinger equation using the displacement operator $\hat{\Dcal}(t) = \hat{\Dcal}_{1}(t)\hat{\Dcal}_{2}(t)$, with $\hat{\Dcal}_{k}(t) = \exp [\xi_{k}(t) \hat{r}_{k}^{\dagger} - \xi_{k}^{\ast}(t) \hat{r}_{k}]$, with $\xi_{k}(t)$ the complex amplitude of the coherent displacement. After this transformation and using the dispersive interaction regime, each resonator mode is driven to a time-dependent coherent state $\ket{\xi_{k}(t)}$, with
\begin{align}
\xi_{k}(t) = \frac{ \varepsilon_{\mathrm{d},k}(t)}{\omega_{\mathrm{r},k} - \omega_{\mathrm{d},k}}e^{-i\omega_{\mathrm{d},k} t + i\phi_{\mathrm{d},k}} . \label{Eq:xi}
\end{align}

In addition, by following this strategy, the atomic evolution is reduced to the dynamics as driven by an effective time-dependent Hamiltonian of the form
\begin{equation}
\hat{H}_{\mathrm{eff}}(t) 
= \hbar \left(  \omega_{0} + \delta_{\mathrm{ds}}\right) \hat{\sigma}_{+}\hat{\sigma}_{-}
-
\hbar \left[\Omega_{\mathrm{res}}(t)\hat{\sigma}_{+} + \Omega_{\mathrm{res}}^{\ast}(t)\hat{\sigma}_{-}\right] , \label{Eq:H_eff}
\end{equation}
where we have already reduced the atom dynamics to the subspace composed of the lowest energy states $\ket{g}$ and $\ket{e}$, valid in the limit of large anharmonicity. The above equation takes into account the frequency shifts, $\delta_{\mathrm{ds},k} = g^{2}_{k}/\Delta_{k}$, promoted by the dispersive cavity-atom interaction in the atom given by $\delta_{\mathrm{ds}} = \delta_{\mathrm{ds},1} + \delta_{\mathrm{ds},2}$. Also, the second term in the above equation is the resulting driving phasor \textit{on the atom}, with complex amplitude $\Omega_{\mathrm{res}}(t)$, due to the coherent driving fields applied to the resonators, which reads 
\begin{equation}
\Omega_{\mathrm{res}}(t) = \sum_{k=1}^{2} \Omega_{0,k}(t) e^{-i\omega_{\mathrm{d},k} t + i\phi_{\mathrm{d},k}} . \label{Eq:OmeRes}
\end{equation}
with $\Omega_{0,k}(t) = g_{k}\varepsilon_{\mathrm{d},k}(t) / (\omega_{\mathrm{r},k} - \omega_{\mathrm{d},k})$ defined as the effective driving field amplitude applied to the $k$-th waveguide. Thus, it is possible to highlight the connection between the Eqs.~\eqref{Eq:H_eff} and~\eqref{Eq:OmeRes} with the Eq.~\eqref{Eq:H_res}, which allows us to conclude that the system considered in Fig.~\SubFig{Fig:Scheme}{a} could be efficiently simulated using the superconducting device in Fig.~\SubFig{Fig:Scheme}{b}. As the main consequence of the Eq.~\eqref{Eq:OmeRes}, again we observe that the atom is driven through an interference-like process as a result of two drives applied to the resonators. Unlike the previous example, this system provides a flexible framework for further discussions due to the number of parameters accessible in the system.

First of all, let us assume the set of frequencies for the resonator modes and qubit as $\omega_{\mathrm{r},1} < \omega_{0} < \omega_{\mathrm{r},2}$, with the particular choice $\omega_{\mathrm{r},1} = \omega_{0} - \Delta_{0}$ and $\omega_{\mathrm{r},2} = \omega_{0} + \Delta_{0}$. These assumptions are not a mere choice, and they are motivated by two main reasons. First, the condition $\omega_{\mathrm{r},1} \neq \omega_{\mathrm{r},2}$ is required to guarantee that the two modes do not effectively interact with each other even in the presence of indirect resonator-resonator couplings~\cite{Hornibrook:12}---similar to crosstalk and parasitic couplings in superconducting qubit systems~\cite{Chow:11,Zhao:22,Santos:23a}---. In fact, even in the case where the resonators are placed far apart, collateral couplings due to the superconducting capacitance network have to be taken into account~\cite{Yan:18,Rasmussen:21,Rosario:23}. Second, when we assume identical resonator-qubit couplings $g_{1}=g_{2}=g$, when $\omega_{\mathrm{r},1}$ and $\omega_{\mathrm{r},2}$ are anti-symmetrically detuned from $\omega_{0}$, the atomic dispersive shift $\delta_{\mathrm{ds}}$ vanishes, as $\delta_{\mathrm{ds},1} = - \delta_{\mathrm{ds},2}$. In this way, the control of the qubit through external drives only needs to take into account the original atomic transition $\omega_{0}$, but it is not a mandatory requirement.

We now consider identical drives applied to each resonator to control the giant atom, with amplitude $\varepsilon_{\mathrm{d}}(t)$ and tuned at atom frequency, i.e., $\omega_{\mathrm{d},k} = \omega_{0}$. In this way, any constructive or destructive interference effect in the resulting phasor in Eq.~\eqref{Eq:OmeRes} will depend on the sign of the effective amplitude $\Omega_{0,k}(t) = (-1)^{k} g\varepsilon_{\mathrm{d}}(t) / \Delta_{0}$, with $k = \{1,2\}$, and the drive's phase matching $\delta \phi = \phi_{\mathrm{d},1} - \phi_{\mathrm{d},2}$. As a result of those sets of assumptions, we obtain $\Omega_{0,1} = -\Omega_{0,2}$, which is a physical situation different from the case of the atom interacting with two lasers (c.f. Fig.~\SubFig{Fig:Scheme}{a}). However, this represents the ideal scenario to establish a second, counterintuitive result of this work: non-overlapping, identical, and \textit{in-phase} waves can \textit{interfere destructively}. This occurs because the interference phenomenon depends on the mode–observer coupling~\cite{Villas-Boas2025}; thus, different couplings yield distinct interference patterns. As an additional outcome of our findings, we introduce here a novel method for controlling interference between waves---namely, by individually modulating the mode–observer coupling.

To confirm the predictions obtained through the approximated analytical results, we implement the numerical exact integration of the Eq.~\eqref{Eq:Dynamics} with the set of assumptions done above. To this end, we use experimentally feasible values of parameters~\cite{Wallraff:04,Schuster:07,Swiadek:24,Tomonaga:25} for the resonators and atom as $\omega_{0} = 5.0\times 2\pi$~GHz, and $g = 80.0\times 2\pi$~MHz, such that the dispersive regime is satisfied with $\Delta_{0} = 2.0\times 2\pi$~GHz. The drive amplitude in each resonator is constant $\varepsilon_{\mathrm{d},k}(t) = \varepsilon_{0} = 100\times 2\pi$~MHz, such that $\Omega_{0,1} = -\Omega_{0,2} = -4.0\times 2\pi$~MHz. In Fig.~\SubFig{Fig:Scheme}{c} we show the behavior of the atom excitation for two different choices of the relative phase $\delta \phi$ between the drives.

When we apply $\pi$-phase shift pulses ($\delta \phi = \pi$) to the system, the atom gets excited through the two independent drives, with full population inversion expected at the excitation time $\tau_{\mathrm{e}} = \pi/4|\Omega_{0,1}|$, estimated from the approximated prediction $\tau_{\mathrm{e}}$ and highlighted in Fig.~\SubFig{Fig:Scheme}{c} with the horizontal dashed line. One concludes that such $\tau_{\mathrm{e}}$ is twice as fast as the inversion time $\tau_{\mathrm{s}}$ required if a single pulse is used to excite the atom, namely, $\tau_{\mathrm{s}} = \pi/2|\Omega_{0,1}|$~\cite{Blais:04,Blais:07,Santos:23a}. On the other hand, using in-phase pulses ($\delta \phi = 0$), the atom is not excited. As a final remark, in Fig.~\SubFig{Fig:Scheme}{c} we show the behavior of the average number of photons in the resonators, showing that in both cases the resonators are weakly populated through the off-resonant external drives. The different behaviors of the atom for each choice of $\delta \phi$ constitute a clear signature of the constructive and destructive interference of the resulting phasor in Eq.~\eqref{Eq:OmeRes}. As the main interpretation of this result, we have shown an interference-like effect between \textit{non-overlapping} driving fields applied to independent and spatially separated resonators.

\emph{Conclusions.--} In this work, we have presented a novel perspective on wave interference by extending its domain beyond the conventional requirement of spatial overlap. By considering a spatially extended atom interacting with two independent and non-overlapping fields, we have shown that interference-like effects can emerge purely through the simultaneous atomic coupling to spatially separated sources. This insight challenges the traditional wave-centric view of interference and highlights the role of the detector’s structure and coherence in shaping observable outcomes. Our theoretical model, supported by a realistic proposal for implementation using superconducting circuits, lays the foundation for new experimental platforms to explore this generalized form of interference.

The non-overlapping wave interference model introduced here can be interpreted as an experimental proposal to test the ideas presented in~\cite{Villas-Boas2025}. One of the central claims of that work is the presence of photons in the dark spots of double-slit experiments---an assertion that is experimentally challenging to verify because, at those points, the photons are not coupled to the detectors and, consequently, cannot be directly measured. In our configuration, the modes are spatially separated, allowing individual measurements of the photons in each mode, even though they interact with the same giant atom. In this way, dark states could be observed by monitoring the excited-state population of the giant atom, while direct measurements on the modes would confirm the presence of photons within them. In addition, the theory considered in our work can be applied to other systems, like superconducting qubits coupled to mechanical acoustic wave resonators~\cite{Von:22,Chou:25,Wollack:22,Von:24}. This would constitute a macroscopic quantum test of our hypothesis in solid-state mechanical resonators, which would allow us to go beyond photonic interference discussed here.

The notion of interference between non-localized fields invites consideration of a broader set of scenarios, including interactions between fields of distinct frequencies coupled asymmetrically to the same atomic transition. Such configurations can lead to either complete scattering or full transparency, depending on the nature of the interference---constructive or destructive---thus enabling tunable control over light–matter interactions. The ability to engineer perfect interference between modes of unequal amplitude by adjusting their coupling strengths further expands the toolkit for manipulating quantum systems. These results not only deepen our understanding of interference in quantum electrodynamics but also pave the way for innovative applications in quantum control, sensing, and the development of atom-transparent devices modulated by spatially disjoint fields. 

\begin{acknowledgments}
ACS acknowledges the financial support from the Comunidad de Madrid through the program Talento 2024 `César Nombela', under Grant No. 2024-T1/COM-31530 (Project SWiQL). C.J.V.-B. acknowledges the support from S\~{a}o Paulo Research Foundation (FAPESP), grant No. 2022/00209-6, and the National Council for Scientific and Technological Development (CNPq), grant No. 311612/2021-0. The authors thank André Cidrim, from DF-UFSCar, for useful discussions during the initial stage of this project.
\end{acknowledgments}


%

\end{document}